\documentclass{article}
\usepackage{spconf,amsmath,graphicx}
\usepackage{booktabs}
\usepackage{array}
\usepackage{tikz}
\usepackage{tikzscale}
\usepackage{tikz-network}
\usepackage{graphics}
\usepackage{graphicx}
\usepackage{pgfplots}
\usepackage{adjustbox}
\usepackage{multirow}
\usepackage{enumitem}
\usepackage{amsfonts,amssymb}
\usepackage{hyperref}
\usepackage{xcolor}

\pgfplotsset{compat=1.11,
    /pgfplots/ybar legend/.style={
    /pgfplots/legend image code/.code={%
       \draw[##1,/tikz/.cd,yshift=-0.25em]
        (0cm,0cm) rectangle (3pt,0.8em);},
   },
}

\newcolumntype{?}{!{\vrule width 1pt}}

\definecolor{bblue}{HTML}{4F81BD}
\definecolor{rred}{HTML}{C0504D}
\definecolor{ggreen}{HTML}{9BBB59}
\definecolor{ppurple}{HTML}{9F4C7C}

\title{Injecting Text and Cross-lingual supervision in few-shot learning from Self-Supervised Models}
\name{Matthew Wiesner$^1$, Desh Raj$^2$, Sanjeev Khudanpur$^{1,2}$}
\address{$^1$Human Language Technology Center of Excellence, Johns Hopkins University, USA\\$^2$Center for Language and Speech Processing, Johns Hopkins University, USA}
\email{\{wiesner,khudanpur\}@jhu.edu, draj@cs.jhu.edu}
\begin{document}
\ninept
\maketitle
\begin{abstract}
   Self-supervised model \emph{pre-training} has recently garnered significant interest, but relatively few efforts have explored using additional resources in \emph{fine-tuning} these models.
   We demonstrate how universal phoneset acoustic models can leverage cross-lingual supervision to improve transfer of pretrained self-supervised representations to new languages. We also show how target-language text can be used to enable and improve fine-tuning with the lattice-free maximum mutual information (LF-MMI) objective. In three low-resource languages these techniques greatly improved few-shot learning performance.
\end{abstract}
\section{Introduction}
\label{sec:intro}

In the last decade, automatic speech recognition (ASR) systems have benefited from supervised training of deep neural networks on transcribed speech at scale. However, since annotated resources for low-resource languages are often scarce or expensive, considerable efforts have been made towards unsupervised and semi-supervised learning for ASR, resulting in methods such as representation learning, pseudo-labeling, local prior matching, and adversarial training~\cite{Kahn2020LibriLightAB,Ling2020DeepCA,Hsu2021SemiSupervisedES,Liu2019AdversarialTO}.
Recently, self-supervised approaches~\cite{Kahn2020SelfTrainingFE, vqvae,cpc,apc,vae,vqwav2vec} have become popular for learning general representations of the input distribution without supervision. In particular, wav2vec2.0 (W2V2)~\cite{baevski2020wav2vec} 
has been used for downstream tasks such as cross-lingual ASR~\cite{conneau2020unsupervised}, speech translation~\cite{Nguyen2020InvestigatingSP}, and speaker and language identification~\cite{Fan2021ExploringW2}. 
\begin{figure}
    \centering
    \includegraphics[trim=0 0 55 0,clip,width=\columnwidth]{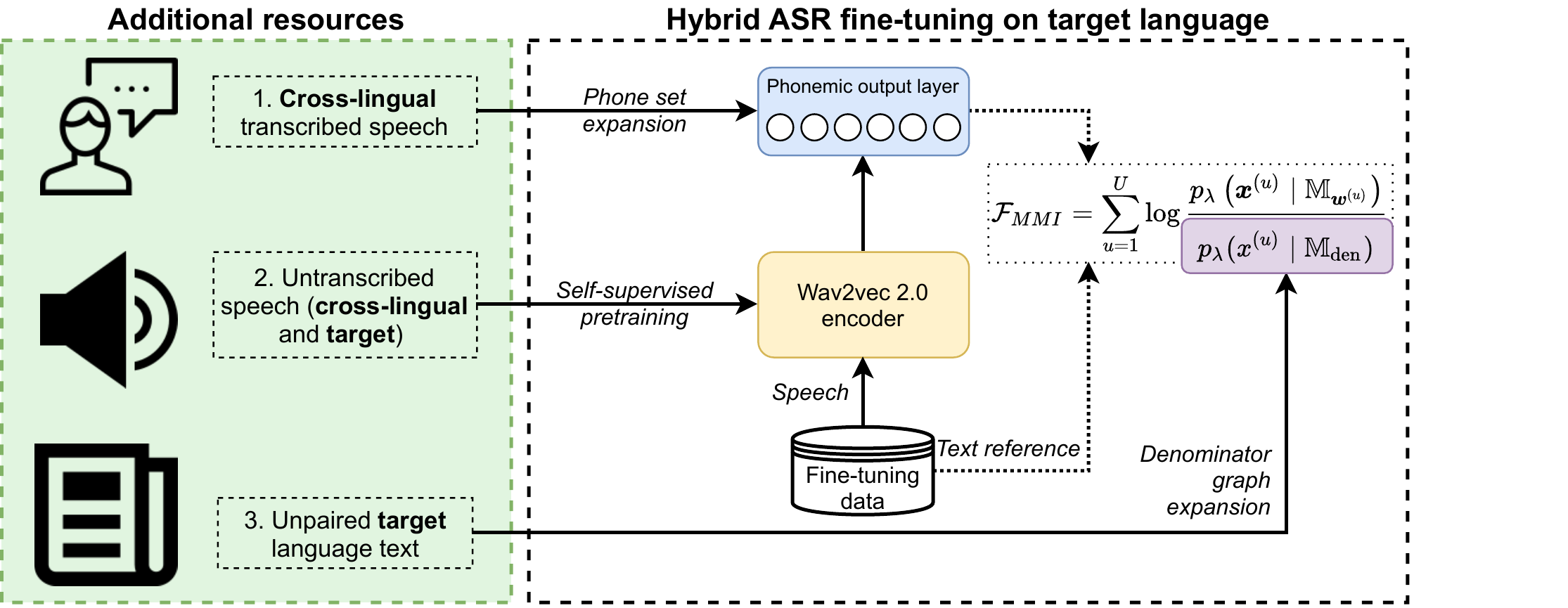}
    \caption{Our proposed integration of cross-lingual transcribed speech and unpaired target language text in LF-MMI-based fine-tuning of large self-supervised models. Cross-lingual supervision is injected using a sequence-trained universal phone recognizer at the output layer. We use unpaired target language text to estimate the denominator graph used for LF-MMI computation.}
    \label{fig:overview}
\vspace{-1em}
\end{figure}

Prior work has investigated improved self-supervised pre-training of models using side information including (i) using a text-to-speech (TTS) module to inject text~\cite{chen2021injecting}, (ii) using teacher models to improve representation learning~\cite{hsu2021hubert}, and (iii) moving average based learning from a fixed supervised teacher~\cite{Zhang2021XLSTCS}. 
However, research on few-shot fine-tuning with these models has received less attention. In this work, we develop ASR models for new languages using \emph{very} small amounts (say, 15 mins.) of transcribed speech. However, we have access to additional resources, including cross-lingual transcribed and untranscribed speech, as well as unpaired target language data. We explore the best strategies to incorporate these in \emph{fine-tuning} large self-supervised pre-trained models. 

\subsection{Related Work}

Most recent research on fine-tuning of self-supervised models is in the ``end-to-end'' framework by learning a linear output layer using the connectionist temporal classification (CTC) objective~\cite{ctc}. In contrast, ``hybrid'' ASR systems which combine acoustic, pronunciation, and language models for decoding with weighted finite state transducers (WFSTs)~\cite{Mohri2002WeightedFT} are popular in industry due to their compositionality and efficiency. We focus on such hybrid ASR models.

Acoustic model transfer in hybrid ASR has used the lattice-free MMI (LF-MMI) objective~\cite{lfmmi} --- in conjunction with methods such as weight transfer, multi-task training, and teacher-student learning~\cite{Ghahremani2017InvestigationOT,Manohar2018ATL} --- for fine-tuning models. One outstanding problem in LF-MMI based model transfer is how best to construct the denominator graph when using only small amounts of domain-matched transcribed speech. This issue was raised in \cite{manohar2017jhu}, but has not been explicitly addressed in a multilingual setting.

Recently, Vyas et al.~\cite{lfmmiwav2vec} studied the application of supervised fine-tuning using LF-MMI on self-supervised acoustic models, and showed that representations learned on English transferred well to low-resource languages such as Swahili. However, there have been no investigations about the use of additional resources, such as cross-lingual supervision or unpaired target language text, in fine-tuning from large self-supervised models. 

We leverage ideas from 
prior work \cite{schultz1997fast, schultz1998multilingual, schultz2001language} which explored language-\emph{independent} approaches for cross-lingual transfer by relying on a universal phoneset, such as the IPA \cite{ladefoged1990revised}, to train phone models by mixing training data from all languages. More recently, the sequence training of universal phoneset models using LF-MMI was explored with various neural architectures~\cite{tong2019investigation, madikeri2020lattice}, while \cite{dalmia2018sequence} explored \emph{cross}-lingual transfer of acoustic models trained using CTC. Our main contributions in this paper are three-fold:
\begin{enumerate}[wide, labelwidth=!, labelindent=0pt]
    \item We propose using a cross-lingual universal phoneset model for few-shot learning. This significantly outperforms the conventional method of fine-tuning by constructing a new output layer.
    
    \item We leverage untranscribed speech by exploring the choice of pre-trained, self-supervised models used in fine-tuning. Experiments show that using matched-domain and matched-language data in pretraining significantly improves downstream performance.
    
    \item We simultaneously leverage unparied, target language text while addressing the question of how to construct the denominator graph in LF-MMI based fine-tuning, using unpaired text to improve the LF-MMI denominator graph phone language model.
\end{enumerate}

\section{Technical Preliminaries}
\label{sec:prelim}

\textbf{The hybrid HMM-DNN framework} for speech recognition follows the Bayes decomposition of the ASR objective. Given a speech input $\textbf{x}^{(u)}$, the most likely sequence $\hat{\mathbf{w}}^{(u)}$ is given as
\begin{align}
\hat{\mathbf{w}}^{(u)} &= \text{arg}\max_{\mathbf{w}} p(\mathbf{w}|\textbf{x}^{(u)}) = \text{arg}\max_{\mathbf{w}} p(\mathbf{x}^{(u)}|\mathbf{w})p(\mathbf{w}) \\
    &= \text{arg}\max_{\mathbf{w},\mathbf{\ell}} \underbrace{p(\mathbf{x}^{(u)}|\mathbf{\ell})}_{H}\underbrace{p(\mathbf{\ell}|\mathbf{w})}_{L}\underbrace{p(\mathbf{w})}_{G},
\end{align}
where each term, $H$, $L$, $G$, are estimated independently. $H$ is a collection of hidden Markov models (HMM), one per (context-dependent) phone, where a neural network models HMM state emission probabilities. Together, the HMMs and neural network are termed the acoustic model. The acoustic model scores can be combined with a pronunciation lexicon, $L$, and language model (grammar), $G$, to produce scores for hypothesis sequences, $\textbf{w}$. One advantage of these models is the ability to optimize each component independently. In this work, this property enables the re-use of acoustic models trained to produce units from a universal phoneset. 

The acoustic model is often trained using a sequence discriminative objective such as maximum mutual information (MMI), which aims to maximize the the point-wise mutual information between the speech signal and reference transcripts:
\begin{equation}
\mathcal{F}_\text{MMI} = \sum_{u \in U} \log \frac{p_{\lambda}(\textbf{x}^{(u)} | \mathbb{M}_w^{(u)})}{p_{\lambda}(\textbf{x}^{(u)}|\mathbb{M}_\text{den})}.
\end{equation}

Here, $\mathbb{M}_w$, called the numerator graph, is a graph representing the set of all alignments and pronunciations corresponding to the ground-truth sequence $w$.   $\mathbb{M}_\text{den}$, called the denominator graph, is ideally, an HMM modeling \emph{all} possible word sequences. The LF-MMI objective~\cite{lfmmi} maximizes a lower-bound on the mutual information \cite{variational_bounds_on_mi} by using un-normalized distributions parameterized by a neural network. The denominator graph is approximated with a 4-gram phone language model. In other words, the LF-MMI objective is \emph{globally} normalized, as opposed to CTC, which assumes a uniform prior between output units and is \emph{locally} normalized. $\lambda$ is the set of all HMM parameters, including the DNN that estimates the probability of HMM states at each time step.

\textbf{Contrastive self-supervised models} aim to learn representations of speech that maximize a lower bound on the mutual information between a chunk of audio $\mathbf{x}_t$ at time $t$ and its surrounding context $\mathbf{c}_t$ using the InfoNCE objective. In wav2vec 2.0~\cite{baevski2020wav2vec}, a convolutional feature extractor generates latent speech representations $\mathbf{z}$ from the waveform, which are then processed by a Transformer to produce contextualized representations $\mathbf{c}$. A quantization module converts the continuous latent vectors $\mathbf{z}$ to discrete representations $\mathbf{q}$. The pre-training objective predicts the correct quantized vector $\mathbf{q}_t$ by contrasting against distractors. The emission probabilities in DNN-HMMs can be replaced by a linear transformation of the outputs $\mathbf{c}$, of self-supervised models. General practice, which we follow, is to freeze the parameters of the feature extractor during fine-tuning.




\section{Method}

In this section, we will describe our method for injecting each of the 3 different ``additional'' resources (shown in Fig.~\ref{fig:overview}) --- cross-lingual transcribed speech, untranscribed speech, and unpaired target language text --- into few-shot learning.

\subsection{Cross-lingual transcribed speech}

We use transcribed cross-lingual speech to train a universal phoneme-sequence model, which we refer to as ``multi-phoneset'' (as opposed to a ``mono-phoneset'' obtained from just the target language). We hypothesize that multi-phoneset models can be used, to an extent, in new languages without transcribed speech. 

\begin{figure}[t]
\centering
\resizebox{\columnwidth}{!}{%
\begin{tikzpicture}
\begin{axis}[
    xlabel=\# Utterances,
    ylabel=WER \%,
    height=3.6cm, width=8cm,
    xmode=log,
    ymin=54.0,
    ymax=100.0,
    xmin=50,
    xmax=10000,
    legend style={at={(1.0,1.0)}},
    x label style={at={(axis description cs:0.5,-0.15)},anchor=north},
    y label style={at={(axis description cs:-0.08,.5)},anchor=south},
    ]
\addplot[smooth,mark=*,blue] plot coordinates {
    (50, 76.1)
    (100, 74.1)
    (200, 70.3)
    (400, 69.7)
    (800, 67.0)
    (1600, 62.8)
    (3200, 61.4)
    (6400,59.8)
    (10000, 56.2)
};
\addlegendentry{Multi-phoneset}
\addplot[densely dotted,mark=square*,red] plot coordinates {
    (50, 100)
    (100, 89.2)
    (200, 80.9)
    (400, 72.8)
    (800, 67.9)
    (1600, 62.0)
    (3200, 59.7)
    (6400,57.7)
    (10000, 54.2)
};
\addlegendentry{Mono-phoneset}

\addplot[dashed,mark=-,black] plot coordinates {
    (50, 59.2)
    (10000, 59.2)
};
\addlegendentry{Scratch}
\end{axis}
\end{tikzpicture}%
}
\caption{Comparison between multi and mono phoneset outputs with increasing training data on WRN models on \texttt{hat}. In the few-shot scenario ($<$ 1k utterances), using a shared, universal phoneset provides significant WER improvements over a monolingual baseline.}
\label{fig:multi_vs_mono}
\vspace{-0.5em}
\end{figure}
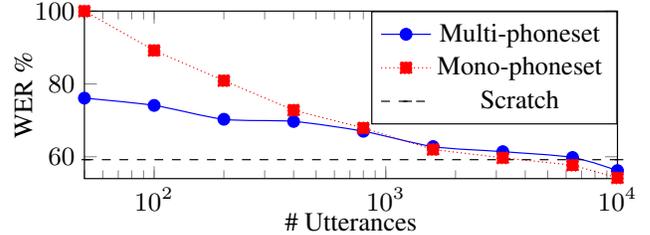

To verify this, we conducted some preliminary experiments on \texttt{hat}, where we analyzed the performance of wide ResNet (WRN) based mono- and multi-phoneset models when fine-tuned on various amounts of data (Fig.~\ref{fig:multi_vs_mono}). We found that in the few-shot scenario ($<$1k utterances), the multi-phoneset model outperformed the conventional approach of retraining a monolingual output layer. Our goal is to replicate this behavior using larger self-supervised models capable of producing better pre-trained representations. 

\subsection{Untranscribed speech}
We explore the value of using matched-domain and matched-language untranscribed speech in self-supervised pre-training. \cite{hsu2021robust} demonstrated the clear value of pre-training on matched acoustic channel untranscribed speech. Furthermore, some of the improvements of self-supervised contrastive learning appear to transfer across language~\cite{kawakami2020learning,riviere2020unsupervised}, and \cite{conneau2020unsupervised} demonstrated that representations trained on large and diverse cross-lingual speech could outperform those trained on small amounts of matched-language speech. Using pre-trained models on speech in new languages results in both mismatch in language and/or acoustic channel (different auditory environments). It is not immediately clear which, if any, of these are more harmful for downstream performance. To investigate these questions, we performed fine-tuning experiments with publicly available W2V2 models pre-trained on 4 different kinds of data:
\begin{enumerate}[wide, labelwidth=!, labelindent=0pt]
\item \textbf{LV60}\footnote{\url{https://hf.co/facebook/wav2vec2-large-lv60}}: Trained on 53.2k hours of English audiobook recordings from Libri-Light~\cite{Kahn2020LibriLightAB}. Few-shot transfer using this model performs remarkable well on LibriSpeech~\cite{Panayotov2015LibrispeechAA}. 
\item \textbf{Large-Robust}\footnote{\url{https://hf.co/facebook/wav2vec2-large-robust}}: Trained on English audiobooks (LibriLight~\cite{Kahn2020LibriLightAB}), conferences (TED-Lium~\cite{Hernandez2018TEDLIUM3T}), and telephone conversations (Switchboard)~\cite{hsu2021robust}.
\item \textbf{VoxPopuli-100k}\footnote{\url{https://hf.co/facebook/wav2vec2-large-100k-voxpopuli}}: Trained on 100k hours of speech in 23 languages from the VoxPopuli corpus ~\cite{wang2021voxpopuli}. The data is obtained from European parliamentary recordings (and only contains European languages).
\item \textbf{XLSR-53}\footnote{\url{https://hf.co/facebook/wav2vec2-large-xlsr-53}}: Trained on 56k hours of speech in 53 languages, including CommonVoice and BABEL corpora. It has previously shown strong cross-lingual transfer performance for ASR and phoneme recognition tasks~\cite{conneau2020unsupervised,Xu2021SimpleAE}.
\end{enumerate}
\begin{table}[t]
\centering
\caption{Previous SOTA WERs (using FLP training) for the languages used for fine-tuning experiments in this paper.}
\label{tab:sota}
\adjustbox{max width=\linewidth}{%
\begin{tabular}{@{}lccc@{}}
\toprule
\textbf{Language} & \textbf{Model} & \textbf{Data (h)} & \textbf{WER} \\ \midrule
Pashto (\texttt{pus}) & CTC-BLSTM + lang. adversarial ~\cite{yi2019language} & 0.2k & 43.4 \\
Haitian (\texttt{hat}) & Multi-lingual CTC-BLSTM~\cite{dalmia2018sequence} & 0.5k & 44.3 \\
Georgian (\texttt{kat}) & XLSR-53 w/ CTC fine-tuning~\cite{conneau2020unsupervised} & 56k & 31.1 \\ \bottomrule
\end{tabular}
}
\vspace{-1.9em}
\end{table}
\vspace{-4mm}
\subsection{Unpaired target language text}
\vspace{-1mm}
Conventionally, unpaired target language text is used to train language models which are then fused with acoustic model outputs during decoding or rescoring. In contrast, we explore how to use these resources during \emph{training} of the acoustic model. Recall from \S\ref{sec:prelim} that the LF-MMI objective, contrary to the CTC objective (often used in fine-tuning self-supervised models), enables us to incorporate a language model in training through the denominator graph $\mathbb{M}_\text{den}$, which can be supplemented with additional unpaired, or cross-lingual text. We conjecture that such denominator graph expansion may lead to better loss computation by estimating a prior which is closer to the true target distribution (as opposed to CTC's uniform prior), particularly in the few-shot scenario.

\section{Experimental Setup}

\subsection{Data}

Prior work in \emph{few-shot} transfer has mainly used clean, read speech.
A motivation of this work was to study how these representations transfer in moderately challenging, conversational, and multilingual speech. The BABEL corpus~\cite{harperdata} was developed to support research in ASR and keyword search in such conditions. The speech data are 8kHz telephone conversations in 25 languages. Each language has two training sets: the full language pack (FLP) (between 40h and 80h); and the limited language pack (LLP) (10h subset of FLP). A 10h development split (\texttt{dev10h}) is also defined for each language. We report performance on \texttt{dev10h} as separate evaluation sets are not publicly available. To study cross-lingual transfer in the \textbf{few-shot} scenario, we created 200-utterance subsets from the LLP set for each language (or 15 min. of speech total from 2-4 speakers).

Pronunciation lexicons containing X-SAMPA \cite{wells1995computer} pronunciations of all words seen in the FLP and \texttt{dev10h} sets are provided. X-SAMPA phonemes are an ASCII representation of the IPA \cite{ladefoged1990revised}, which was constructed to be a cross-lingual finite symbol set representing all attested phonemes in the world. This label-set enables training universal phoneset models.

We use 21 languages (XL-21) as our cross-lingual supervision, and report performance of all our systems on held-out languages, Pashto (\texttt{pus}), Haitian (\texttt{hat}), and Georgian (\texttt{kat}). Importantly, the FLP audio from these languages are \emph{included} in the XLSR-53 training data. In Table~\ref{tab:sota}, we report the state-of-the-art WERs on these languages using various methods from literature. We used a 1h subset of the Haitian \texttt{dev10h} for early stopping and hyper-parameter tuning.

\begin{table}[t]
\centering
\caption{Effect of cross-lingual supervision on downstream \textbf{few-shot} WER performance, using the 15-min train subset. Column 2 shows the amount of (labeled / unlabeled) data in pre-training.}
\label{tab:xlingual}
\adjustbox{max width=\linewidth}{%
\begin{tabular}{@{}llccc@{}}
\toprule
\textbf{Model} & \textbf{Pre-training (h)} & \texttt{pus} & \texttt{hat} & \texttt{kat} \\
\midrule
XL-21 WRN Mono & 0 / 1.2k & 86.2 & 81.5 & 89.2 \\
XL-21 WRN Multi & 0 / 1.2k & 75.6 & 72.9 & 73.1 \\
XL-21 Wav2Vec 2.0 (random) & 0 / 1.2k  & 77.2 & 77.9 & 76.0 \\
XLSR-53  & 56k / 0 & 75.9 & 74.7 & 81.5\\
XLSR-53 + XL-21  & 56k / 1.2k & 71.0  & 67.8 & 64.5\\
XLSR-53 (frozen) + XL-21 Multi & 56k / 1.2k & \textbf{63.6}  & \textbf{58.9} & \textbf{58.2}\\
\bottomrule
\end{tabular}
}
\vspace{-1em}
\end{table}

\begin{table*}[t]
\centering
\caption{Effect of pre-training with untranscribed cross-lingual data on the downstream fine-tuning WER. $^\dag$The WERs can be reduced to 38.8, 36.9, and 32.2, respectively, if we decode with an expanded vocabulary with no OOVs.}
\label{tab:w2v2}
\adjustbox{max width=\linewidth}{%
\begin{tabular}{@{}lccccccccccc@{}}
\toprule
\multirow{2}{*}{\textbf{Model}} & \multicolumn{2}{c}{\textbf{Pre-training (hours)}} & \multicolumn{3}{c}{\textbf{Few-shot ($\sim$15m)}} & \multicolumn{3}{c}{\textbf{LLP ($\sim$10h)}} & \multicolumn{3}{c}{\textbf{FLP ($\sim$80h)}} \\
 \cmidrule(r{4pt}){2-3} \cmidrule(r{4pt}){4-6} \cmidrule(lr){7-9} \cmidrule(l{4pt}){10-12}
 & \textbf{Labeled} & \textbf{Unlabeled} & \multicolumn{1}{l}{\texttt{pus}} & \multicolumn{1}{l}{\texttt{hat}} & \multicolumn{1}{l}{\texttt{kat}} & \multicolumn{1}{l}{\texttt{pus}} & \multicolumn{1}{l}{\texttt{hat}} & \multicolumn{1}{l}{\texttt{kat}} & \multicolumn{1}{l}{\texttt{pus}} & \multicolumn{1}{l}{\texttt{hat}} & \multicolumn{1}{l}{\texttt{kat}} \\
\midrule
Random WRN & 0 & 0 & 93.2 & 95.4 & 95.4 & 61.1 & 57.7 & 57.6 & 50.1 & 46.6 & 49.5 \\
XL-21 WRN Mono & 1.2k & 0 & 86.2 & 81.5 & 89.2 & 56.6 & 52.3 & 55.1 & 44.7 & 43.3 & 46.2 \\
\midrule
LV60 & 0 & 60k & 81.4 & 84.2 & 86.0 & 50.3 & 50.3 & 50.5 & 42.8 & 40.8 & 43.1 \\
Large-Robust & 0 & 63k & 81.2 & 80.1 & 93.0 & 49.5 & 49.1 & 48.7 & 41.8 & 40.4 & 42.2 \\
VoxPopuli-100k & 0 & 100k & 80.3 & 77.2 & 85.6 & 48.9 & 48.0 & 47.9 & 41.1 & 39.3 & 41.5 \\
XLSR-53 & \multirow{1}{*}{0} & \multirow{1}{*}{56k} & \textbf{75.9} & \textbf{74.7} & \textbf{81.5} & \textbf{45.5} & \textbf{45.3} & \textbf{45.1} & \textbf{39.4}$^\dag$ & \textbf{37.7}$^\dag$ & \textbf{40.0}$^\dag$ \\
\bottomrule
\end{tabular}%
}
\vspace{-1em}
\end{table*}

\subsection{Training details}

We used 2 types of acoustic models: 28-10 wide ResNets (WRN) \cite{wrn}, and the W2V2-based models and architectures. All monolingual and cross-lingual supervised baselines are reported using the WRN model; for example, XL-21 WRN refers to the WRN model pre-trained on the 21 BABEL languages with supervision.

We conducted fine-tuning experiments on the few-shot, LLP, and FLP subsets, where the training alignments for the few-shot and LLP were obtained by training GMM-HMMs on the LLP subset, and those for FLP by training on the FLP data, respectively. We use the cross-lingual GMM-HMMs, occasionally remapping missing phonemes in the target language to a similar phoneme seen in training to create the targets for fine-tuning with the "Multi" models. We trained all acoustic models with PyTorch~\cite{Paszke2019PyTorchAI} and used Kaldi~\cite{Povey2011TheKS} for decoding. We used Pychain for the LF-MMI implementation \cite{pychain}. All decoding was performed with a 3-gram language model trained on the FLP transcripts in srilm \cite{srilm}. This conforms with a common scenario where larger amounts of unpaired text are more readily available than transcribed speech and is different from denominator graph expansion, which we used during \emph{training}. Our implementation and recipes are publicly available\footnote{\url{https://github.com/m-wiesner/nnet_pytorch}}. The denominator graph phone-LMs for the ``multi'' models combined the cross-lingual and target language training data, weighting the target-language data by a factor of 10 as in \cite{manohar2017jhu}.

\section{Results and Discussion}
\subsection{Effect of phone set expansion}
Table \ref{tab:xlingual} shows the few-shot transfer of multilingual baselines in the three target languages. We find that transferring the supervised cross-lingual representation from WRN models (row-2) works better than reconstructing the output layer (row-1). Training the cross-lingual supervised model with the W2V2 neural architecture (row-3) performed slightly worse than the WRN baseline. Self-supervised vs. supervised pre-training (rows 3 v. 4) made little difference in the few-shot scenario. There were likely not enough data to retrain a output transformations of the self-supervised representations.

We compare 2 approaches that combine the supervised, sequence-level universal phoneset model, with the pretrained self-supervised models. In the XLSR-53 + XL-21 (row-5), we performed two sequential fine-tuning steps: first on the cross-lingual representations, and then on the target language few-shot supervisions. We also try freezing the XLSR-53 network when fine-tuning on the XL-21 data (row-6). This was the best-performing approach, almost matching the \emph{supervised} models trained on 10h of audio! We suspect that freezing the W2V2 model prevents catastrophic forgetting during cross-lingual fine-tuning, while simultaneously enabling us to learn a universal phoneset model. Few-shot learning by transferring a universal output layer worked significantly better than reconstructing the output layer using small amounts of speech as in \cite{baevski2020wav2vec, conneau2020unsupervised}.

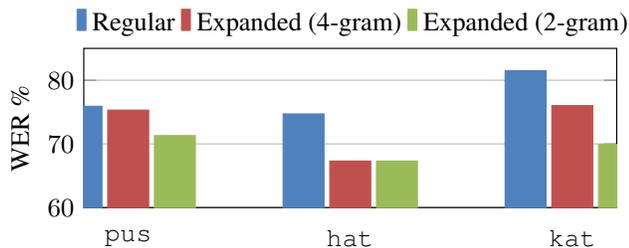
\begin{figure}[t]
\centering
\resizebox{\columnwidth}{!}{%
\begin{tikzpicture}
    \begin{axis}[
        width  = 8cm,
        height = 3.5cm,
        major x tick style = transparent,
        ybar,
        bar width=14pt,
        ymajorgrids = true,
        ylabel = {WER \%},
        symbolic x coords={\texttt{pus},\texttt{hat},\texttt{kat}},
        xtick = data,
        scaled y ticks = false,
        legend cell align=left,
        legend style={draw=none,at={(1.05,1.3)}},
        legend columns=3,
        ymin=60,
        ymax=85,
    ]
        \addplot[style={bblue,fill=bblue}]
            coordinates {(\texttt{pus}, 75.9) (\texttt{hat},74.7) (\texttt{kat},81.5)};

        \addplot[style={rred,fill=rred}]
            coordinates {(\texttt{pus}, 75.3) (\texttt{hat},67.3) (\texttt{kat},76.0)};

        \addplot[style={ggreen,fill=ggreen}]
            coordinates {(\texttt{pus}, 71.3) (\texttt{hat},67.3) (\texttt{kat},69.9)};
        
        \legend{Regular,Expanded (4-gram),Expanded (2-gram)}
    \end{axis}
\end{tikzpicture}

%
}
\caption{Effect of denominator graph expansion (regular vs. expanded) for LF-MMI based few-shot cross-lingual transfer of the XLSR-53 model.}
\label{fig:den_graph}
\vspace{-1em}
\end{figure}

\subsection{Analysis of pre-trained W2V2 models} Self-supervised pre-training outperformed XL-21 based supervised pre-training in all scenarios, as shown in Table~\ref{tab:w2v2}, but we found a clear hierarchy among pretrained models. \textbf{(i) Matched unlabeled data is king} -- XLSR-53 was the best performing base-model and improved over a cross-lingual supervised WRN baseline by 10-15\% relative across the board. It was trained on the smallest amount of, but \emph{matched} acoustic and language, data. The LV60 model is trained on a similar amount of data, but only clean English audio from LibriLight \cite{Kahn2020LibriLightAB}. It was by far the worst performing model. \textbf{(ii) Linguistic diversity, quantity of data, and matched acoustic conditions are important}. For instance, the VoxPopuli-100k model, like the LV60 model, is trained on clean wide-band speech. However, presumably due to the quantity, and multilingual (but not matched language) nature, of training data, it was the second best performing model. Finally, the Robust model, despite training on much less, and purely \emph{monolingual} English speech, performed comparably to the VoxPopuli model, likely thanks to the telephone speech (similar to BABEL) that it was exposed to in training. Increasing the amount of fine-tuning data, reduced the difference in performance among models.

\subsection{Effect of denominator graph expansion}

\begin{table}[t]
    \centering
    \caption{Effect of $n$-gram order and unpaired text weight, $\alpha$, on WER in LF-MMI few-shot learning of Haitian ASR models.}
\begin{tabular}{@{}llllllll@{}}
\toprule
\textbf{Weight} ($\alpha$) & \textbf{n=0} & \textbf{n=1} & \textbf{n = 2} & \textbf{n = 3} & \textbf{n = 4} \\ 
\midrule
$\alpha=0.0$ & 81.9 & 66.2 & 65.2 & 68.2 & 74.7  \\
$\alpha=0.1$ & 81.9 & 66.1 & \textbf{65.0} & 66.3 & 67.6 \\
$\alpha=0.2$ & 81.9 & 66.3 & 65.1 & 66.1 & 66.8\\
$\alpha=0.5$ & 81.9 & 65.9 & 65.2 & 66.3 & 67.2 \\ \bottomrule
\end{tabular}

    \label{tab:order_v_weight}
    \vspace{-5mm}
\end{table}

Using additional, unpaired text to train the denominator graph phone-LM reduces overfitting.  Fig.~\ref{fig:den_graph}, shows that using the resulting "expanded" LF-MMI denominator graphs improves the WER across all languages in the few shot setting. However, using lower-order $n$, in the phone-LM $n$-gram model would result in less over-fitting. We analyzed the trade-off between modeling-power and unpaired text weight in the Haitian few-shot scenario with XLSR-53 (Table~\ref{tab:order_v_weight}).

The best preforming systems use $n=2$, demonstrating the benefit of using LF-MMI in fine-tuning as opposed to CTC, which corresponds to the $n=0$ case with a different HMM topology. We also found that including the unpaired text with a small weight improved performance across all, but especially higher-order $n$. Using additional text enables the use of higher-order n-gram models even in low-resource scenarios. We also see that the improvements from the universal phoneset model cannot be attributed solely to the inclusion of cross-lingual phone sequences when training the denominator graph phone-LM.   

\section{Conclusion}

In this work, we have shown how cross-lingual supervision can be used during fine-tuning in few-shot learning to significantly improve transfer of self-supervised models. We confirmed the importance of using matched-domain speech in pre-training, and we demonstrated how to incorporate external text in the phone-LM of the LF-MMI denominator graph in order to help guide fine-tuning. These techniques resulted in significantly improved few-shot performance in three low-resource test conditions. 

\footnotesize
\bibliographystyle{IEEEbib}
\bibliography{refs}

\end{document}